\documentclass[fleqn,usenatbib]{mnras}

\usepackage{newtxtext,newtxmath}

\usepackage[T1]{fontenc}

\DeclareRobustCommand{\VAN}[3]{#2}
\let\VANthebibliography\thebibliography
\def\thebibliography{\DeclareRobustCommand{\VAN}[3]{##3}\VANthebibliography}

\input{anc/gaiadr3_sb1_commands}
\usepackage{graphicx}	
\usepackage{amsmath}	
\usepackage{siunitx}
\usepackage{booktabs}
\usepackage{dcolumn}
\usepackage[flushleft]{threeparttable}

\title[Compact objects among Gaia DR3 SB1s]{A search for compact object companions to high mass function single-lined spectroscopic binaries in Gaia DR3}

\author[T. Jayasinghe et al.]{T. Jayasinghe,$^{1,2,*,\thanks{E-mail: jayasinghearachchilage.1@osu.edu}}$,
D.M. Rowan$^{1,2,*,\thanks{E-mail: rowan.90@osu.edu}}$,
Todd A. Thompson$^{1,2,3}$,
C. S. Kochanek$^{1,2}$,
K. Z. Stanek$^{1,2}$
\\
$^{1}$Department of Astronomy, The Ohio State University, 140 West 18th Avenue, Columbus, OH, 43210, USA\\
$^{2}$Center for Cosmology and Astroparticle Physics, The Ohio State University, 191 W. Woodruff Avenue, Columbus, OH, 43210, USA\\
$^{3}$Department of Physics, The Ohio State University, Columbus, Ohio, 43210, USA\\
$^{*}$These authors contributed equally to this work.\\
}
\date{Accepted XXX. Received YYY; in original form ZZZ}

\pubyear{2022}

\begin{document}
\label{firstpage}
\pagerange{\pageref{firstpage}--\pageref{lastpage}}
\maketitle

\begin{abstract}
\textit{Gaia} DR3 provides $>181,000$ radial velocity solutions for single-lined spectroscopic binaries (SB1s) which can be used to search for non-interacting compact object$+$star binary candidates by selecting systems with large mass functions. We selected \nTotal{} such systems and identified \nGraded{} systems with good RV solutions in DR3. We used light curves from ASAS-SN and \textit{TESS} to identify and remove \nEBsTotal{} eclipsing binaries to produce a catalog of \nVetted{} compact object+star candidates, including \nELLsTotal{} ellipsoidal variables. The positions of these candidates on \textit{Gaia} and 2MASS CMDs suggest that many of these systems are binaries with luminous companions. We compared the periods and eccentricities of detached eclipsing binaries in \textit{Gaia} DR3 and ASAS-SN, and found that ${\sim}11\%$ and ${\sim}60\%$ of the binaries had different periods and eccentricities. We also compared RV solutions for 311 binaries in both \textit{Gaia} DR3 and the Ninth Catalog of Spectroscopic Binary Orbits (SB9), and found similar results. We do not identify any strong candidates for non-interacting compact object+star binaries.
\end{abstract}

\begin{keywords}
stars: black holes -- (stars:) binaries: spectroscopic
\end{keywords}



\section{Introduction}

The mass distribution of neutron stars and stellar mass black holes is closely tied to the evolution of massive stars, including their deaths  (e.g., \citealt{Pejcha2015, Sukhbold2016, Woosley2020}). The relationship between the pre-supernova mass of the massive star and the type of compact remnant left behind is complex, and depends on the chemical composition of the star, mass-loss rates, supernova explosion physics and binary interactions (e.g., \citealt{Sukhbold2016, Patton2021}). While the fates of individual massive stars are hard to ascertain, we do know that dead massive stars leave behind a plethora of compact remnants in our Galaxy -- there are predicted to be about ${\sim}10^8$ stellar mass BHs and ${\sim}10^9$ neutron stars in the Milky Way (e.g., \citealt{Brown1994}).

A well characterized, unbiased sample of neutron stars and black holes is necessary to better understand massive stars. However, this is a challenging task because the vast majority of compact objects are electromagnetically dark. To date, most mass measurements for neutron stars and black holes come from pulsar and accreting binary systems selected from radio, X-ray, and gamma-ray surveys (see, for e.g., \citealt{Champion2008,Liu2006,Ozel2010,Farr2011,Strader2015}), and from the LIGO/Virgo detections of merging systems (see, for e.g., \citealt{TheLIGOScientificCollaboration2021, Abbott2016,Abbott2017NS}). The populations of BHs observed as X-ray binaries and gravitational wave (GW) mergers are both heavily biased samples. In X-ray binaries, the companion must either fill its Roche lobe or have a modest separation and a strong stellar wind. Compact objects discovered through GW observations come from the small fraction of surviving binaries that are on very tight orbits leading to a merger. These interacting systems are, however, a small minority of compact object binaries, and the far larger population of non-interacting systems is essentially unexplored \citep{Tanaka2002, Wiktorowicz2019}. While non-interacting binaries are harder to find, they must be discovered and characterized in order to fully understand the numbers, properties, formation mechanisms, and evolutionary pathways of the interacting systems.

Rapid advances in time-domain astronomy \citep{Shappee2014,Kochanek2017,Jayasinghe2018,Bellm2019,Tonry2018,Lindegren2021,GaiaCollaboration2022} provide promising pathways to discovering of non-interacting compact objects. For example, \citet{Chawla2021} estimated that ${\sim}30-300$ non-interacting black holes are detectable in binaries around luminous companions using \textit{Gaia} astrometry. Similarly, \citet{Shao2019} used binary population synthesis models to estimate that there are thousands of non-interacting black holes in the Milky Way, with hundreds of these systems having luminous companions with $G<20$~mag. In addition to astrometry, targeted searches combining high-cadence photometry and sparsely sampled radial velocities from wide-field time-domain surveys are a promising method to discover more systems (e.g., \citealt{Trimble1969,Thompson2019,Zheng2019,Rowan2021}). 

The discovery and confirmation of non-interacting compact objects is a challenging endeavour. As a result, only a handful of convincing non-interacting compact objects other than pulsars have been discovered thus far. Three non-interacting BH candidates have been discovered in globular clusters: one by \citet{Giesers2018} in NGC 3201 (minimum black hole mass  $M_{\rm BH} = 4.36 \pm0.41$\,M$_\odot$), and two by \citet{Giesers2019} in NGC 3201 ($M_{\rm BH}\sin(i) = 7.68 \pm0.50$\,M$_\odot$ and $M_{\rm BH}\sin(i) = 4.4 \pm2.8$\,M$_\odot$). These globular cluster systems, if they indeed contain black holes, likely have formation mechanisms that are very different from those of field black hole binaries because the high stellar densities allow formation mechanisms which do not operate for field stars. A single convincing non-interacting BH candidate has been found in the field. \citet{Thompson2019} discovered a low-mass ($M_{\rm BH}\simeq3.3_{-0.7}^{+2.8}~M_\odot$) non-interacting black hole candidate in the field on a circular orbit at $\rm P_{\rm orb}\sim83\,d$ around a spotted giant star. 

However, searches for non-interacting compact objects have also yielded numerous false positives. The binary LB-1 was initially thought to host an massive stellar black hole ($M_{\rm BH}\simeq68_{-3}^{+11}~M_\odot$, \citealt{Liu2019}), but was later found to have a much less massive companion that was not necessarily a compact object (see, for e.g., \citealt{Shenar2020,Irrgang2020,Abdul-Masih2020,El-Badry2020a}). The naked-eye system HR 6819 was claimed to be a triple system with a detached black hole with $M_{\rm BH} = 6.3 \pm0.7~M_\odot$ \citep{Rivinius2020}, but was later found to be a binary system with a rapidly rotating Be star and a slowly rotating B star \citep{El-Badry2020b,Bodensteiner2020}. Recently, NGC 1850 BH1 was claimed to be a binary displaying ellipsoidal variability in the LMC with $M_{\rm BH} = 11.1 _{-2.4}^{+2.1}~M_\odot$ \citep{Saracino2021}, but was later argued to be a stripped B star binary \citep{El-Badry2021}. Another example of a BH imposter was the system NGC 2004 \#115, claimed to be a triple system consisting of a Be star on a tertiary orbit and an inner binary of a B star and a ${\simeq}25~M_\odot$ black hole \citep{Lennon2021}. \citet{El-Badry2021NGC2004} later argued that the orbital inclination was underestimated by assuming tidal synchronization, and that the companion to the B star was more likely a ${\sim}2-3~M_\odot$ main sequence star. \citet{Jayasinghe2021} identified the nearby, nearly edge-on $\rm P_{\rm orb}=59.9$~d circular binary V723~Mon as a candidate for a compact object---star binary. \citet{ElBadry2022} later showed that V723 Mon is better explained by a stripped red giant in a binary around a massive (${\sim}2.8M_\odot$), rapidly rotating subgiant. A common theme to these cases is an overestimate of the mass of the observed star based on its luminosity and the assumption of single star evolution for a binary where mass transfer has greatly reduced the mass of the more luminous star.

\textit{Gaia} DR3 \citep{GaiaCollaboration2022,Babusiaux2022,Katz2022} provides a catalog of 181,529 single-lined spectroscopic binaries (SB1). However, it only provides the orbital solutions--- the individual RV measurements will only be released in \textit{Gaia} DR4. Nonetheless, these SB1 RV solutions are a useful starting point to search for non-interacting compact object+star binaries in the Milky Way. We describe the selection of high mass function binaries from the \textit{Gaia} DR3 SB1 catalog, the vetting process that we used to discard false positives in Section \ref{candidateselectionvet} and compare the RV solutions for binaries in common between \textit{Gaia} DR3 and the \cite{Rowan2022} catalog of detached eclipsing binaries or the SB9 catalog of spectroscopic binaries \citep{Pourbaix2004}. In Section \ref{section:results}, we discuss the vetted candidates. We present a summary of our results in Section \ref{section:summary}.

\section{Candidate Selection and Vetting using Gaia DR3} \label{candidateselectionvet}

We describe the selection of single-lined spectroscopic binaries with high mass functions from Gaia DR3 and our vetting procedure to produce a catalog for the selection and follow-up of candidate non-interacting compact object binaries.

\subsection{Selecting high mass function binaries from Gaia DR3}

\begin{figure*}
    \centering
    \includegraphics[width=0.9\linewidth]{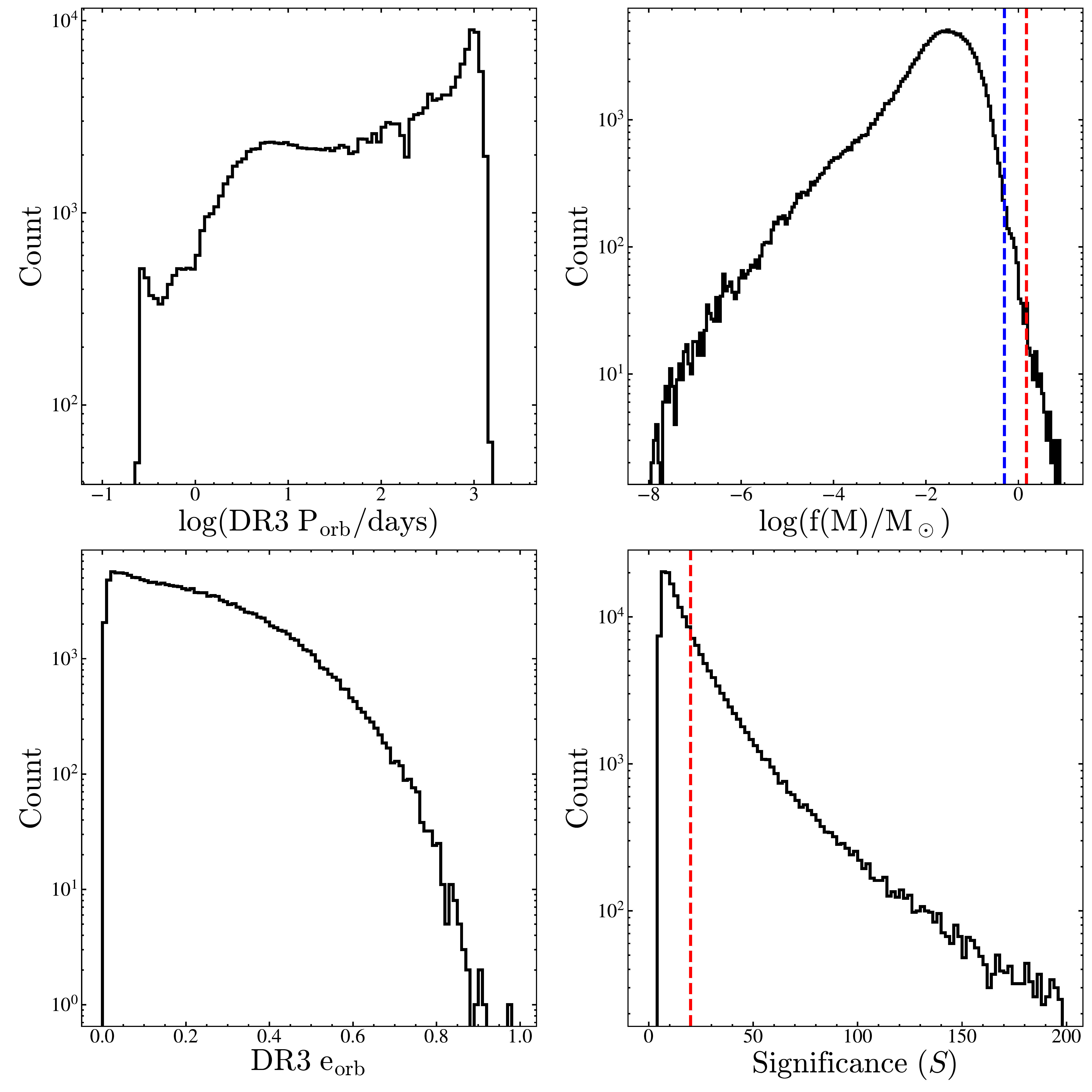}
    \caption{Distributions of the ${\sim}181,500$ \textit{Gaia} DR3 SB1 systems in period, $f(M)$, eccentricity and the significance of the RV solution. In the mass function panel, the red and blue dashed lines show the selection limits for evolved and main sequence stars respectively. The dashed line in the significance panel shows the minimum significance we consider.} 
    \label{fig:sb1stats}
\end{figure*}

Figure \ref{fig:sb1stats} shows the distribution of the \textit{Gaia} DR3 SB1 systems in orbital period, $f(M)$, eccentricity and the \verb"significance" ($S$) of the RV solution. The median binary mass function is $f(M)\simeq~0.018~M_\odot$. Only ${\sim}0.07\%$ of the SB1s have $f(M)>1.5~M_\odot$. The distribution of orbital periods is more complicated, and it reflects the sampling of the \textit{Gaia} survey as well as the true period distribution. However, the RV solutions are skewed towards longer periods--- ${\sim}53\%$ of the DR3 SB1 solutions have an orbital period longer than 100 days. Most systems have modest orbital eccentricities, with a median eccentricity of $e_{\rm orb}\simeq0.19$. Finally, ${\sim}40\%$ (${\sim}3\%$) of the SB1s have significance $S>20$ ($S>100$). For SB1s, the significance is the ratio of the RV semi-amplitude to its uncertainty. The median number of RV epochs used in DR3 is 23, but the RV solutions can have a minimum of 10 epochs or a maximum of 225 epochs, with ${\sim}1.4\%$ of the SB1s having more than 50 individual RV epochs.

\begin{figure*}
    \centering
    \includegraphics[width=1.0\linewidth]{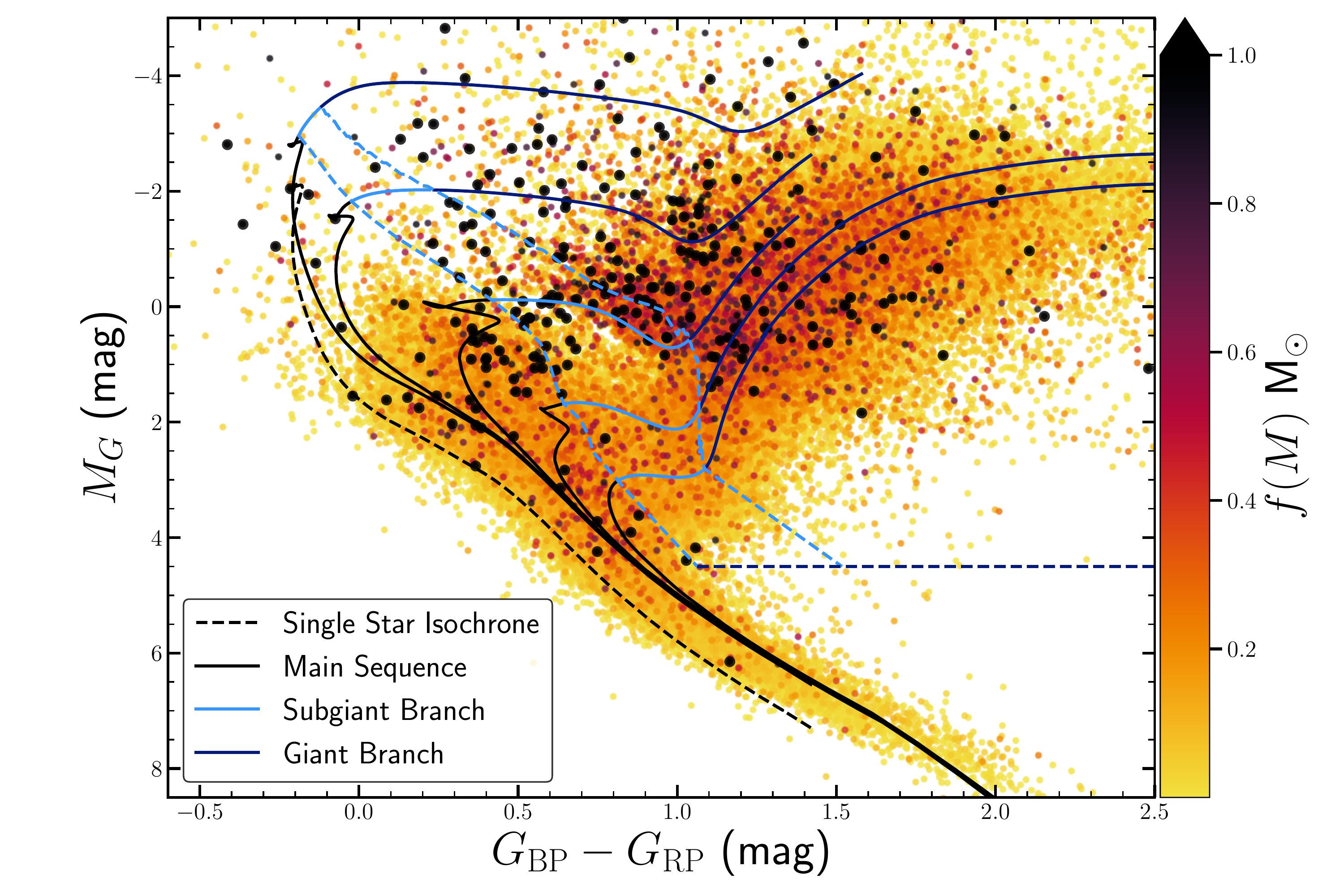}
    \caption{Gaia color-magnitude diagram for the ${\sim}181,000$ \textit{Gaia} DR3 SB1 systems colored by their mass function, $f(M)$. Targets with $f(M)>1.0$ are shown as larger black points. Solar metallicity isochrones for equal mass binaries are shown as the solid lines for ages of $10^8$ to $10^{10}$ in intervals of 0.5 in dex, and divisions of the CMD from \citet{Rowan2022}. The black dashed line shows a single star isochrone. The points are shaded by the binary mass function.}
    \label{fig:cmdsb1all}
\end{figure*}

Figure \ref{fig:cmdsb1all} shows the \textit{Gaia} DR3 CMD for the ${\sim}181,000$ SB1s. We use distances from \citet{Bailer-Jones2021} and compute extinctions with the {\tt mwdust} 3-dimensional `Combined19' dust map \citep{Bovy2016, Drimmel2003, Marshall2006, Green2019}. We use Table 3 of \citet{Wang2019} to convert the $A_V$ extinctions from {\tt mwdust} to $A_G$ and $E(G_{\rm{BP}}-G_{\rm{RP}})$. We use isochrones from MESA Isochrones \& Stellar Tracks \citep[MIST,][]{Choi2016, Dotter2016} to determine the evolutionary state of the candidates, following the prescription in \citet{Rowan2022}. We interpolate over Solar metallicity MIST isochrones ranging in age from $10^8$ to $10^{10}$ years in intervals of $0.1$ in dex to define the boundary between main sequence stars and subgiants. To represent binary star isochrones in Figure \ref{fig:cmdsb1all}, we double the flux in each band to represent an equal mass binary. The end of the subgiant branch is defined at the point when the radius is $R=1.5 R_{\rm{TAMS}}$, where $R_{\rm{TAMS}}$ is the radius at the terminal age main sequence. We set the maximum absolute magnitude limit for the subgiant/giant branch to be $M_G=4.5$~mag. Of the ${\sim}181,000$ SB1s, ${\sim}95,000$, ${\sim}16,000$, and ${\sim}71,000$ are consistent with giants, sub-giants and main-sequence stars, respectively.

We select candidate SB1 systems based on their mass function 
\begin{equation}
    f(M) = \frac{P_{\rm orb}K^3(1-e^2)^{3/2}}{2 \pi G}= \frac{M_{\rm comp}^3 \sin^3i}{(M_{\rm *}+M_{\rm comp})^2},
\end{equation} in the \textit{Gaia} DR3 SB1  catalog. The binary mass function $f(M)$ is the minimum mass ($M_{\rm comp}$) of a companion in a spectroscopic binary in the limit where the mass of the observed star is $M_{\rm *}=0$ so the SB1 systems with high mass functions are good non-interacting compact object binary candidates. The best candidates for non-interacting black hole binaries are those with $f(M)>3~M_\odot$, as the minimum mass exceeds the maximum mass of a neutron star. We selected any system with $f(M)>1.5~M_\odot$ and main sequence systems with $f(M)>0.5~M_\odot$. The limit of $f(M)>1.5~M_\odot$ is to focus on NS and BH candidates. We use a higher threshold for the evolved stars because
it is very difficult to estimate the stellar masses of evolved stars, while it is relatively easy to do so for main sequence stars.  This yields \nTotal{} systems before any cuts on the significance.

\begin{figure}
    \centering
    \includegraphics[width=\linewidth]{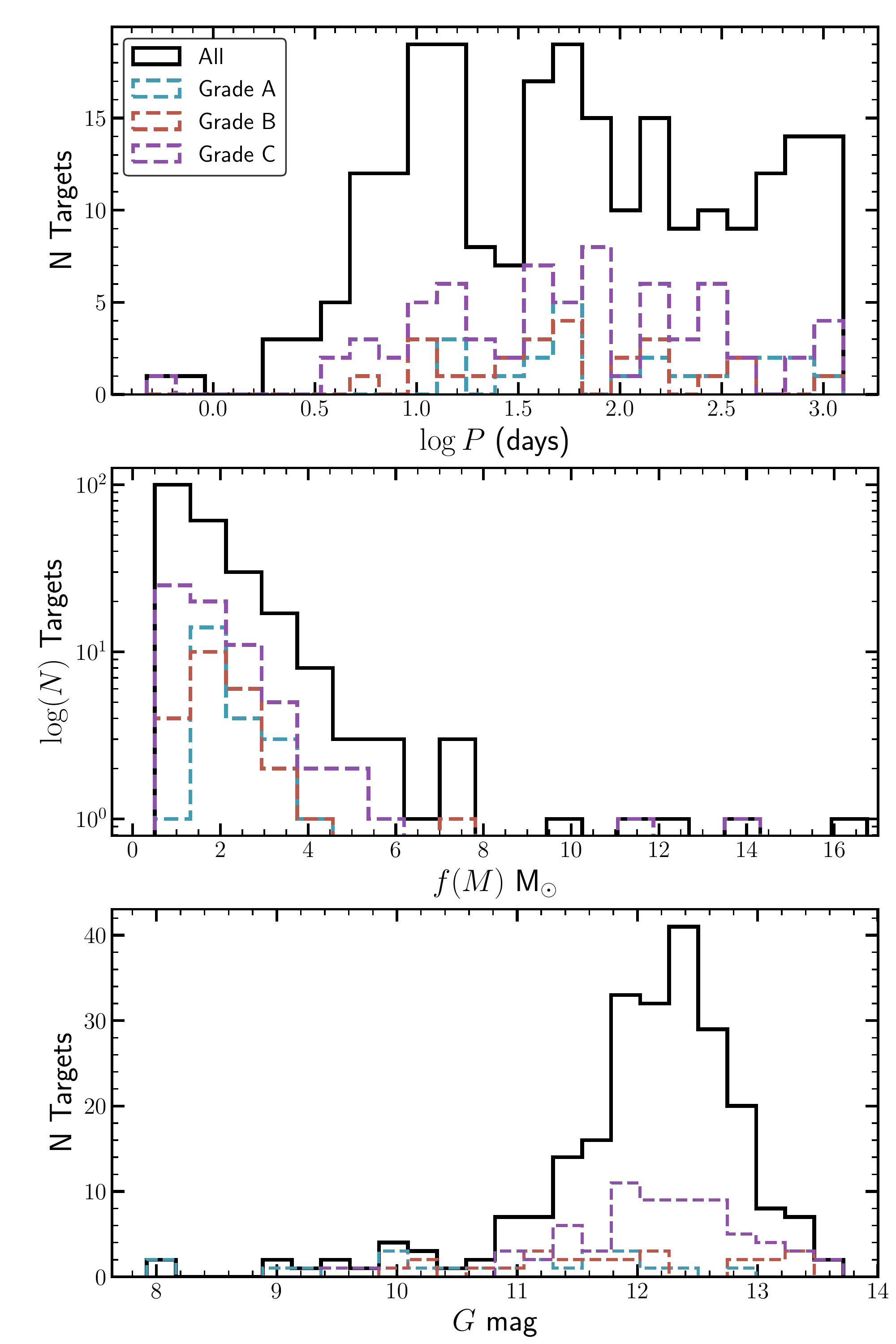}
    \caption{Distributions of the orbital period, mass function, and $G$ magnitude for the high mass function binaries.}
    \label{fig:sb1hist}
\end{figure}

Without the individual RVs for these systems, it is impossible to check the orbital solutions without new spectroscopic observations. For example, without adequate coverage of the maxima and minima in the RV curve, the semi-amplitude ($K$) and eccentricity ($e_{\rm orb}$) might be poorly constrained. In order to limit the false positives that may emerge from poor RV solutions in DR3, we use the \verb"significance" ($S$) parameter in the SB1 catalog to sort our candidates into 4 categories. The best orbital solutions with $S\geq100$ are assigned a grade of `A', those with $50 \leq S< 100$ are assigned a grade of 'B' and those with $20 \leq S< 50$ are assigned a grade of 'C'. The SB1 systems with $S<20$ are assigned a grade of 'D' and are not considered further. The complete list of SB1s is available as supplementary information. There are \nGradeA{}, \nGradeB{} and \nGradeC{} SB1s with grades of A, B and C, respectively. There were 46 additional Grade D systems with $f(M)>1.5~M_\odot$. Figure \ref{fig:sb1hist} shows the distribution of period, $f(M)$ and the DR3 $G$-band magnitude for these systems. The median mass function is $f(M)\simeq2.1~M_\odot$. There are 20 SB1 systems with $f(M)>3.0~M_\odot$. Not surprisingly given the properties of \textit{Gaia} DR3, all of these high $f(M)$ SB1s are relatively bright ($G<14$~mag), and therefore suitable for ground-based radial velocity (RV) and other follow-up observations.

\subsection{Candidate Vetting} \label{sec:lcs}

Here we describe the vetting process used to identify false positives for compact object--stellar binary systems. These include eclipsing binaries (EBs) and systems where the \textit{Gaia} RV solutions are unlikely to be correct. Well-sampled light curves are needed to identify EBs. While  \textit{Gaia} DR3 comes with individual epoch photometry for ${\sim}10$ million variable stars \citep{Eyer2022}, not all of the SB1 systems have \textit{Gaia} light curves and the \textit{Gaia} light curves are generally sparse. Of the \nGraded{} A/B/C systems, only 45 had \textit{Gaia} light curves in DR3. We instead use light curves from the All-Sky Automated Survey for SuperNovae (ASAS-SN; \citealt{Shappee2014,Kochanek2017,Jayasinghe2018}, Hart et al. in prep) and the Transiting Exoplanet Survey Satellite (TESS; \citealt{Ricker2015}) to identify EBs, ellipsoidal variables (ELLs), and systems where the \textit{Gaia} period is inconsistent with the photometric period. We start by cross-matching our catalog with the ASAS-SN $V$- and $g$-band catalogs of variable stars \citep{Jayasinghe2020, Christy2022} and find that a total of \nASASSNTotal{} targets (Grade A: \nASASSNgradeA{}, Grade B: \nASASSNgradeB{}, Grade C: \nASASSNgradeC{}) have been identified as photometric variables with classifications of \ASASSNVarTypes{}. For the remaining targets, we compute $g$-band light curves using the ASAS-SN Sky Patrol \citep{Kochanek2017}. We also inspect the \textit{TESS} light curves from the SPOC \citep{Caldwell2020} and QLP \citep{Huang2020a, Huang2020b, Kunimoto2021} pipelines. Out of the \nGraded{} grade A-C targets, \nTESS{} have \textit{TESS} light curves available for at least one sector. For both of these \textit{TESS} pipelines we use the ``raw'' light curves rather than the detrended light curves because the detrending can remove real stellar variability.

\begin{figure*}
    \centering
    \includegraphics[width=\linewidth]{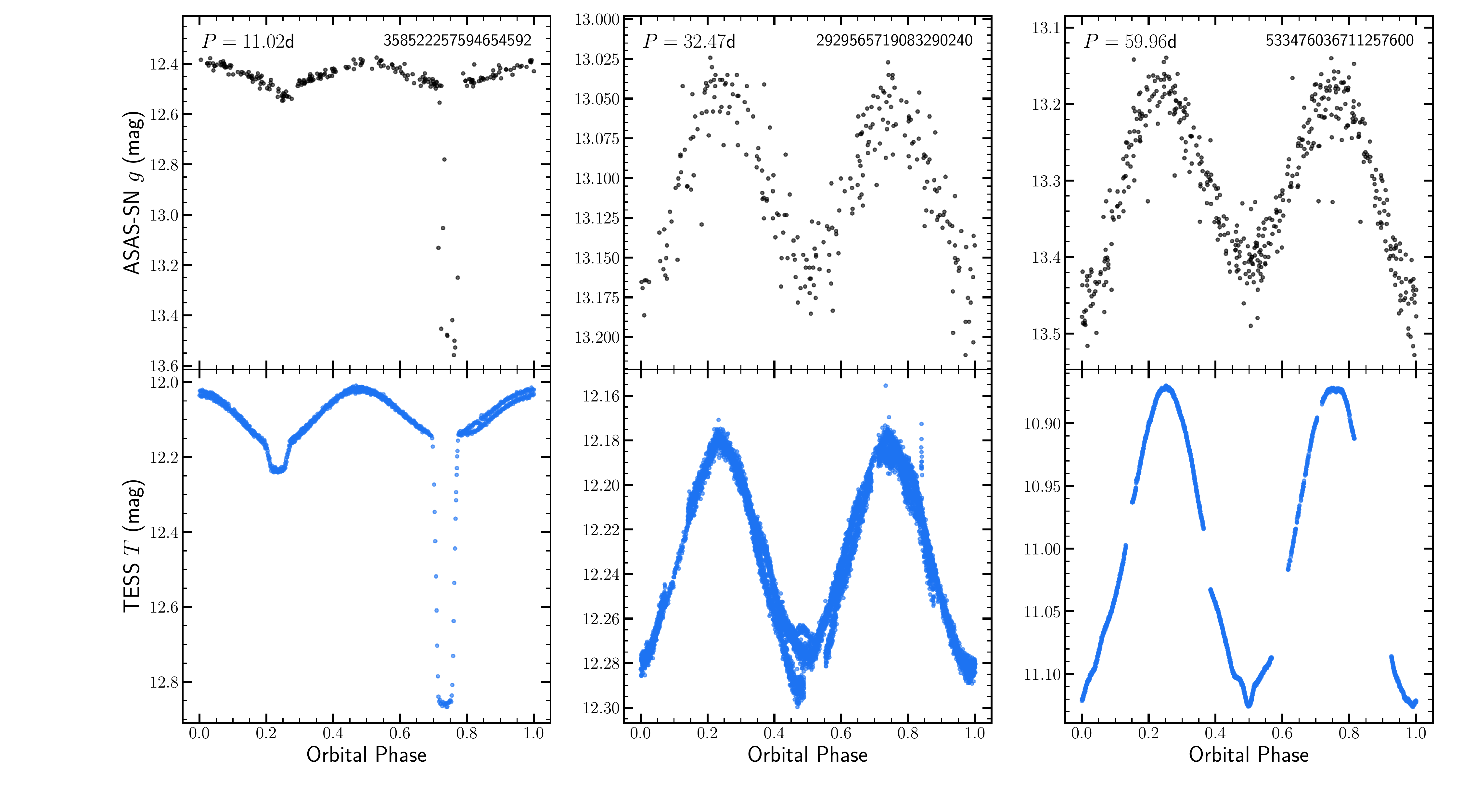}
    \caption{Examples of an eclipsing binary, an ellipsoidal variable, and an eclipsing ellipsoidal variable identified during candidate vetting. The top row shows the ASAS-SN $g$-band light curves and the bottom row shows TESS $T$ light curves. For the eclipsing binary, the orbit has been phased such that the primary eclipse occurs at phase 0.75. The ELL and ELL+ECL system have orbital phase defined such that the deeper minima occurs at phase 0.}
    \label{fig:lc_panel}
\end{figure*}

In total, we identify \nEBsTotal{} eclipsing binaries (Grade A: \nEBsGradeA{}, Grade B: \nEBsGradeB{}, Grade C: \nEBsGradeC{}) that should not be further considered as candidates for hosting a compact object companion. We also identify \nELLsTotal{} ellipsoidal variables (Grade A: \nELLsGradeA{}, Grade B: \nELLsGradeB{}, Grade C: \nELLsGradeC{}). Figure \ref{fig:lc_panel} shows examples of an eclipsing binary, ellipsoidal, and ellipsoidal+eclipsing systems. We find that the \textit{TESS} light curves are especially effective at identifying eclipsing ELL systems that may be missed in the ASAS-SN photometry because the higher photometric precision of the \textit{TESS} data makes it more sensitive to shallow eclipses from a lower-luminosity companion star.

\begin{figure*}
    \centering
    \includegraphics[width=0.9\linewidth]{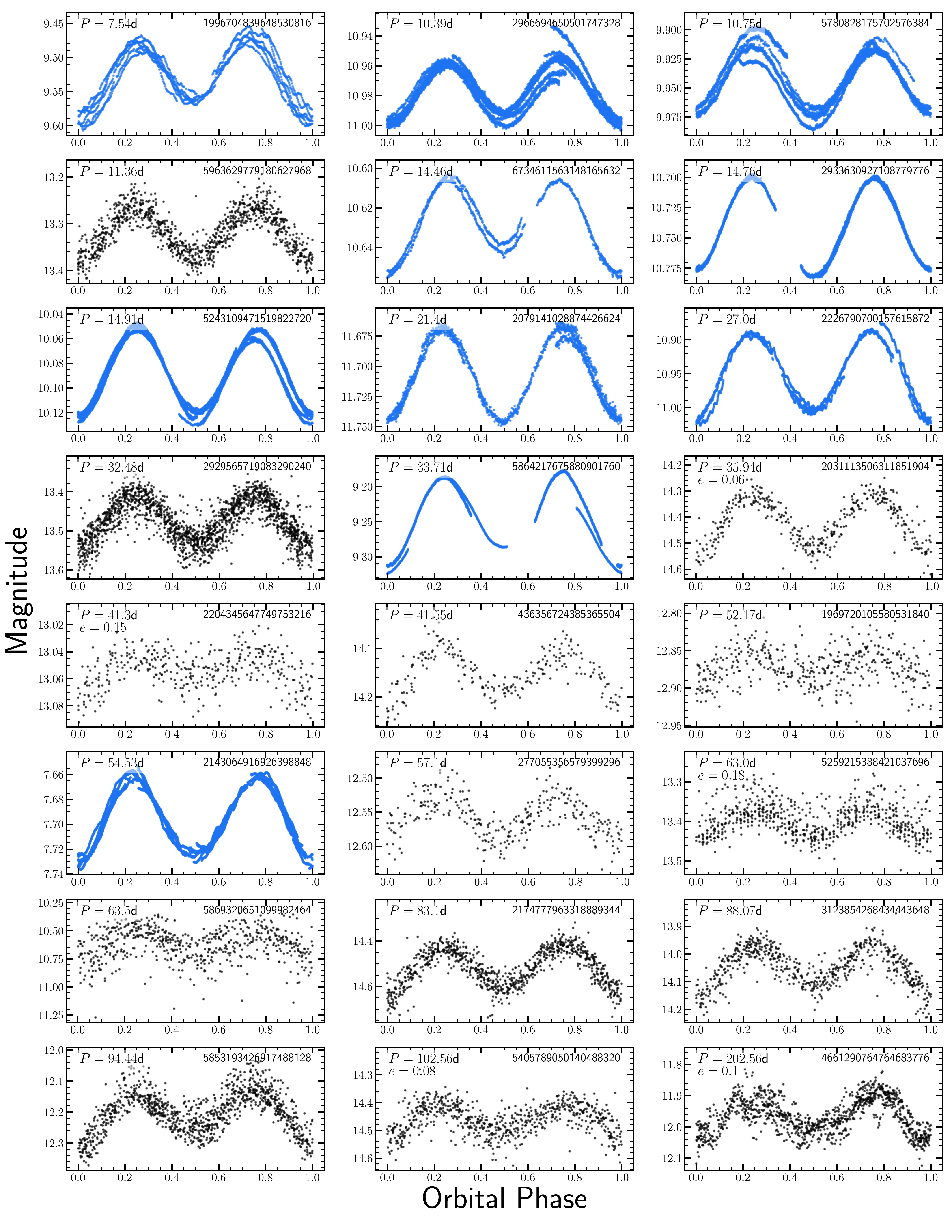}
    \caption{Examples of ellipsoidal variables identified in Section \ref{sec:lcs} from the ASAS-SN and TESS light curves. ASAS-SN light curves are shown in black and TESS light curves are shown in blue. The light curves have all been phased such that the deeper minimum occurs at phase zero. The photometric period is consistent with the Gaia RV period for all but four targets. These targets are removed from further consideration and not included here. The photometric period is given in the upper left of each panel, and the eccentricity from the Gaia SB1 solution is also shown for systems where $e>0.05$. The Gaia source ID is labeled in the upper right of each panel.}
    \label{fig:ell_panel}
\end{figure*}

For the systems with ellipsoidal modulations we also verify that the photometric period is consistent with the Gaia RV period. We use a Generalized Lomb-Scargle \citep{Lomb1976, Scargle1982} to identify the photometric period for targets not included in the ASAS-SN Variable Stars Database. Out of the \nELLsTotal{}, only 4 (\fracELLsWrongPeriod{}\%) have photometric periods that are significantly different ($>10\%$) from the periods given in the Gaia SB1 solution:
\begin{itemize}
    \item 6021285355771958528: $P_{\rm{Gaia}} = 269.51$~d, $P_{\rm{LC}} = 5.71$~d
    \item 5868858821883779328: $P_{\rm{Gaia}} = 46.31$~d, $P_{\rm{LC}} = 34.63 $~d
    \item 878555832642451968: $P_{\rm{Gaia}} = 837.10$~d, $P_{\rm{LC}} = 8.39$~d
    \item 5543340020666175488: $P_{\rm{Gaia}} = 135.75$~d, $P_{\rm{LC}} = 20.91$~d
\end{itemize}

Without access to the individual epochs of Gaia RVs, we are unable to independently fit the RV curve and obtain the correct mass function, so these systems are removed from further consideration. Similarly, the presence of ellipsoidal variability can be used to evaluate the eccentricity reported in the SB1 fit. All of the targets identified as ellipsoidal variables have light curves consistent with zero eccentricity, yet \nELLsHighEcc{} systems have $e>0.05$ from the Gaia SB1 fit. Figure \ref{fig:ell_panel} shows some of the grade A-C ellipsoidal variables that have consistent photometric and spectroscopic periods. 
Out of the \nEBsTotal{} eclipsing binaries identified during the vetting, 13 (\fracEBsWrongPeriod{}\%) have photometric periods that differ by $>10\%$ from the \textit{Gaia} SB1 solution.

\begin{figure*}
    \centering
    \includegraphics[width=0.6\linewidth]{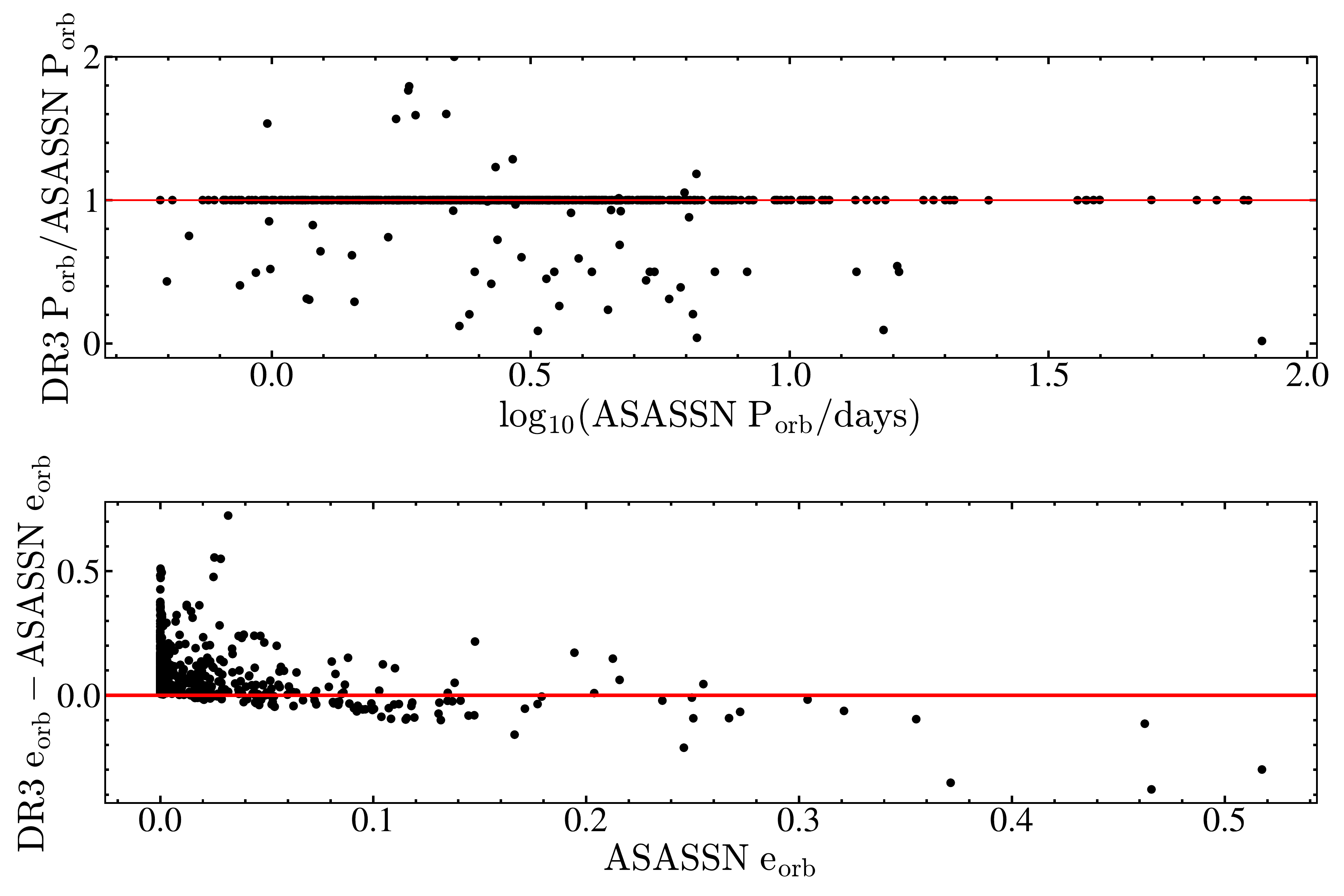}
    \includegraphics[width=0.6\linewidth]{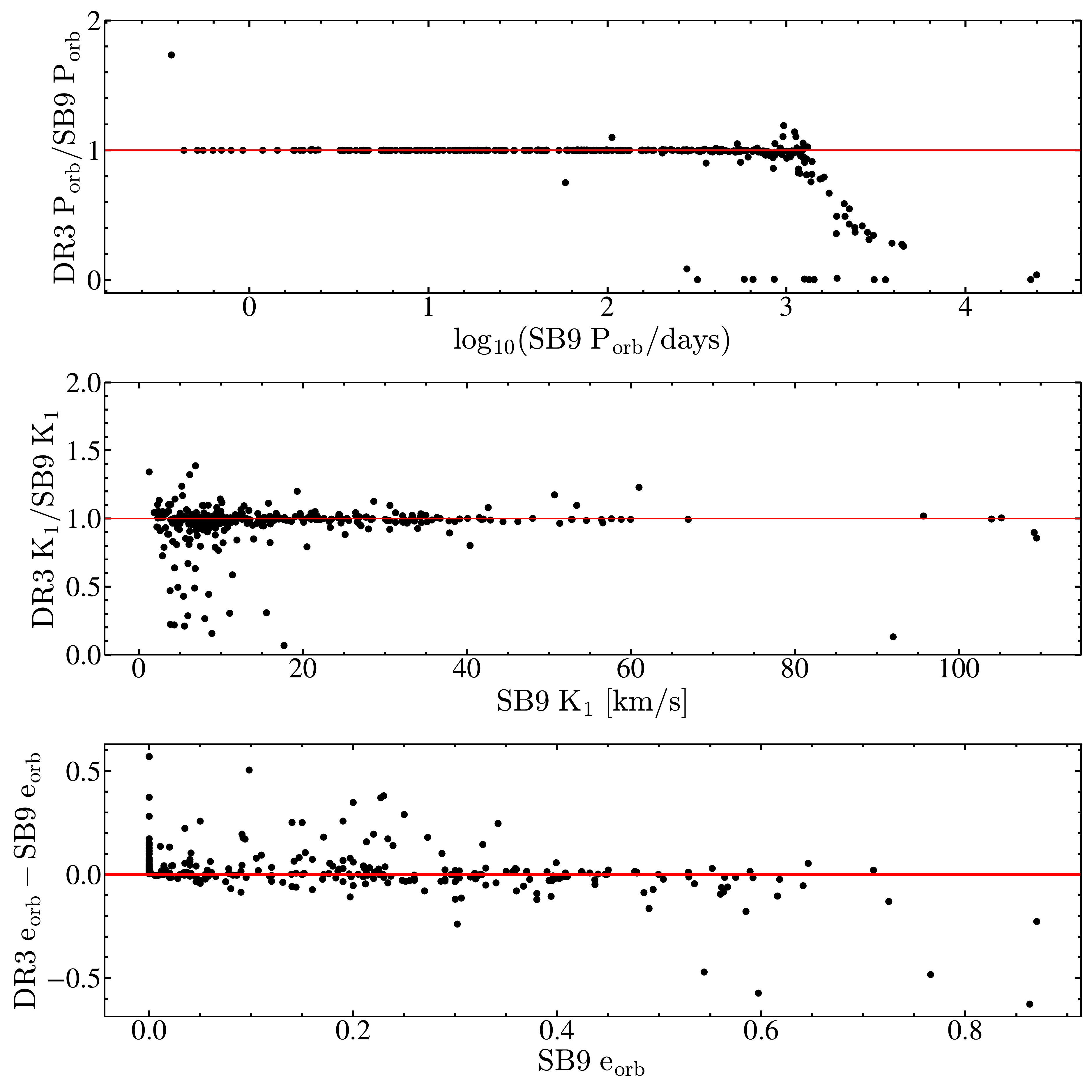}
    
    \caption{The top two panels compare the \textit{Gaia} DR3 and ASAS-SN eclipsing binary periods (upper) and
    eccentricities (lower). The bottom three panels compare the \textit{Gaia} DR3 and SB9 spectroscopic binary
    periods (upper), RV semi-amplitudes (middle) and eccentricities (lower).  The red lines correspond to a one-to-one agreement between both catalogs. Aside from the very long period systems, the agreement in period and
    semi-amplitude is generally good, but there is significant scatter in the eccentricities.}
    \label{fig:asassn_ebs}
\end{figure*}

These period differences, where we would generally expect them to be the same since the photometric variability is also dominated by the brighter star, suggested doing a broader examination of the SB1 solutions. We cross matched the SB1
catalog to the \cite{Rowan2022} catalog of detached eclipsing binaries and the SB9 \citep{Pourbaix2004} catalog of spectroscopic binaries. We identified 596 spectroscopic binaries in common between \textit{Gaia} DR3 and the ASAS-SN detached binaries. Figure \ref{fig:asassn_ebs} shows comparisons of the orbital periods and eccentricities for these detached binaries. Of the 596 systems, ${\sim}89\%$ had periods that agreed to within $\pm10\%$ and ${\sim}40\%$ had eccentricities that agreed to within 0.05. The median fractional error in the DR3 period and the median error in the eccentricity for these detached binaries is ${\sim}0.005\%$ and ${\sim}0.028$ respectively. If we only select the 245 systems that have $S>20$ in \textit{Gaia} DR3, ${\sim}94\%$, and ${\sim}59\%$ had periods, and eccentricities that agreed to within these levels. While most of the binaries have periods that agree, the \textit{Gaia} DR3 eccentricities for many circular systems are overestimated.

We identified 311 spectroscopic binaries in common between \textit{Gaia} DR3 and SB9. Of the 311 systems, 12 systems (${\sim}4\%$) were cataloged as double-lined spectroscopic binaries (SB2) in the SB9 catalog and 245 of these systems (${\sim}79\%$) have $S>20$ in \textit{Gaia} DR3. The median \textit{Gaia} DR3 \textit{G}-band magnitude for these SB1s is ${\sim}8.2$~mag.  Figure \ref{fig:asassn_ebs} also shows comparisons of the orbital periods, RV semi-amplitudes and eccentricities for these SB1 systems. The median fractional error in the DR3 period and the median error in the eccentricity for these systems is ${\sim}0.1\%$ and ${\sim}0.016$ respectively. Of the 311 systems, ${\sim}85\%$, ${\sim}80\%$ had periods and semi-amplitudes that agreed to within $\pm10\%$ and ${\sim}70\%$ had eccentricities that agreed to within 0.05. If we only select the 245 systems that have $S>20$ in \textit{Gaia} DR3, ${\sim}94\%$, ${\sim}90\%$ and ${\sim}81\%$ had periods, semi-amplitudes and eccentricities that agreed to within these levels. 
These comparisons support the use of $S>20$ to select for good RV solutions. The agreement in period and semi-amplitude
is generally very good for orbital periods shorter than ${\sim}1000$~days.  The level of agreement on the eccentricity is
significantly worse, which is not surprising given the typical numbers of RV measurements going into the orbital solutions.

\section{Discussion} \label{section:results}

\begin{table*}
    \caption{The vetted catalog of Gaia DR3 SB1s for Grades A and B. Grade C targets are shown in Table \ref{tab:candidate_tableC}. Targets are sorted by $f(M)$. The `remarks' column labels the systems identified here as ELLs, the classification given in the ASAS-SN $V$- and $g$-band variable stars catalogs, and systems with HI features near 860~nm.}
    \label{tab:candidate_tableAB}
    \begin{threeparttable}
        \sisetup{table-auto-round, group-digits=false}
        \begin{tabular}{l r S[table-format=2.2] r S[table-format=3.2] S[table-format=1.2] S[table-format=3.4] S[table-format=3.2] S[table-format=2.2] r}
\toprule
 {Gaia DR3 Source ID} &     {TIC} &     {$G$} &  {N RVs} &  {Significance} &  {RV Ecc} &  {RV Period} &    {$K_1$} &  {$f(M)$} &     {Remarks} \\
{} & {} & {(mag)} & {} & {} & {} & {(d)} & {(km/s)} & {M$_\odot$} & {}\\ 
\midrule
\large \textbf{Grade A:}\\
   206292746724589824 &  65919311 & 11.567327 &       15 &      134.814390 &  0.027971 &   347.001862 &  50.101911 &  4.516898 &            SR \\
  5869320651099982464 & 443204265 &  9.878837 &       13 &      257.181240 &  0.014177 &    63.923918 &  77.526747 &  3.085621 &           ELL \\
  5864217675880901760 & 319384609 &  9.307944 &       23 &      276.200960 &  0.006531 &    32.957533 &  94.372047 &  2.870202 &           ELL \\
  5853193426917488128 & 397891772 & 11.231160 &       58 &      187.938600 &  0.027227 &    94.546816 &  64.570534 &  2.634646 &       ELL, SR \\
  2021374066702077312 & 113100455 & 10.040213 &       16 &      249.071850 &  0.007008 &    44.413183 &  80.354996 &  2.387668 &            HI \\
  2008602689332951808 & 433877616 & 11.561862 &       19 &      176.193630 &  0.087776 &   746.780507 &  29.567067 &  1.977127 &               \\
   426648861352371328 & 256101165 & 11.765608 &       30 &      198.133480 &  0.008147 &   240.418458 &  42.286516 &  1.883587 &               \\
   277055356579399296 &   9133993 & 11.498751 &       19 &      139.798610 &  0.013096 &    57.082604 &  68.054182 &  1.863854 &           ELL \\
  5541400855806539392 & 182220071 & 12.323505 &       28 &      120.971080 &  0.076998 &   680.690124 &  29.852769 &  1.859881 &               \\
  6734611563148165632 & 405124480 & 10.940159 &       10 &      106.440285 &  0.008308 &    14.344372 & 104.398021 &  1.691098 &     ELL, EB22 \\
  5352456452682656256 & 457864395 & 12.029145 &       27 &      126.030106 &  0.020771 &    59.850214 &  64.053421 &  1.628797 &            HI \\
   510691269572063232 &  54722608 & 11.950563 &       25 &      111.722220 &  0.278042 &   401.907835 &  34.902305 &  1.569386 &               \\
  5243109471519822720 & 371686230 & 10.451161 &       22 &      170.075780 &  0.022729 &    14.913679 & 100.370249 &  1.561416 & ELL, EB22, HI \\
  2143064916926398848 & 243276811 &  7.970002 &       20 &      331.717400 &  0.004292 &    54.130368 &  65.181678 &  1.553320 &           ELL \\
  4061400381769067392 & 198141489 &  7.919531 &       12 &      343.417570 &  0.005649 &   619.989392 &  28.622871 &  1.506463 &               \\
  5722942457613931264 & 386429851 &  9.942932 &       16 &      114.958830 &  0.024770 &    38.099937 &  65.575140 &  1.112153 &               \\
\bottomrule
\midrule
\large \textbf{Grade B:}\\
   527155253604491392 & 444463982 & 13.241814 &       18 &       56.404236 &  0.039641 &   149.155050 &  58.748022 &  3.126430 &              \\
  5934549483335296384 &   8390898 & 13.336455 &       23 &       66.629875 &  0.011894 &    39.459315 &  83.878903 &  2.412520 &              \\
  1969720105580531840 & 273409875 & 11.867422 &       20 &       71.996254 &  0.029184 &    52.245971 &  75.034792 &  2.284237 &          ELL \\
   436356724385365504 & 117261905 & 13.116592 &       20 &       74.995186 &  0.046568 &    41.667586 &  79.763055 &  2.183961 &     ELL, YSO \\
  4116115997318757120 & 351790283 & 10.605717 &       11 &       80.141230 &  0.219198 &    41.298015 &  78.581979 &  1.928752 &              \\
  2226790700157615872 & 376968598 & 11.620328 &       20 &       58.101974 &  0.029358 &    26.896792 &  88.313918 &  1.917263 & ELL, YSO, HI \\
  5405789050140488320 & 432646925 & 13.239762 &       25 &       50.698112 &  0.075597 &   102.598820 &  56.378678 &  1.888899 &      ELL, SR \\
   461251282546634496 & 251656921 & 12.900666 &       22 &       78.366120 &  0.012467 &   457.299371 &  33.544210 &  1.788164 &              \\
  6054973670507378944 & 450908863 & 12.814929 &       16 &       83.004320 &  0.014411 &   127.955769 &  51.231189 &  1.782310 &              \\
  2079141028874426624 & 268608966 & 12.101652 &       19 &       68.671630 &  0.005746 &    21.581811 &  88.101534 &  1.529228 &          ELL \\
  6019286920341931520 & 458320995 & 11.687070 &       27 &       53.460700 &  0.048750 &    47.562497 &  67.450289 &  1.507028 &              \\
  2933630927108779776 &  79033600 & 11.067193 &       20 &       68.974550 &  0.021142 &    14.717528 &  92.918559 &  1.222569 &    ELL, EB22 \\
  2966694650501747328 &  33520660 & 11.203405 &       31 &       79.496640 &  0.019488 &    10.397987 & 102.373299 &  1.155272 &    ELL, EB22 \\
  5780828175702576384 & 324759352 & 10.235976 &       19 &       80.753044 &  0.048828 &    10.749564 &  84.845983 &  0.677879 &          ELL \\
  3102189152022476288 &  36344278 & 11.962961 &       10 &       53.590000 &  0.152794 &     6.085486 & 103.485478 &  0.674477 &              \\
\bottomrule
\end{tabular}

        \begin{tablenotes}
                \item The remark EB22 means the system was included in \citet{El-Badry2022}.
        \end{tablenotes}
    \end{threeparttable}
\end{table*}

\begin{table*}
    \caption{Same as \ref{tab:candidate_tableAB} but for Grade C targets.}
    \label{tab:candidate_tableC}
    \begin{threeparttable}
        \sisetup{table-auto-round, group-digits=false}
        \begin{tabular}{l r S[table-format=2.2] r S[table-format=3.2] S[table-format=1.2] S[table-format=3.4] S[table-format=3.2] S[table-format=2.2] r}
\toprule
 {Gaia DR3 Source ID} &     {TIC} &     {$G$} &  {N RVs} &  {Significance} &  {RV Ecc} &  {RV Period} &    {$K_1$} &  {$f(M)$} &      {Remarks} \\
{} & {} & {(mag)} & {} & {} & {} & {(d)} & {(km/s)} & {M$_\odot$} & {}\\ 
\midrule
\large \textbf{Grade C:}\\
  4661290764764683776 &  31181424 & 11.770099 &       13 &       29.410261 &  0.095333 &   204.929896 &  86.675141 & 13.639446 &            ELL \\
  5259215388421037696 & 274794828 & 12.212911 &       27 &       42.088013 &  0.183672 &    62.706640 & 123.079353 & 11.507172 &            ELL \\
  2006840790676091776 & 388220893 & 11.183457 &       13 &       28.334784 &  0.203377 &     4.073262 & 239.758434 &  5.460109 &                \\
   442992311418593664 & 117801951 &  9.551256 &       20 &       21.781998 &  0.413849 &   216.530553 &  65.532625 &  4.763938 &                \\
  2174777963318889344 & 260523268 & 13.064586 &       23 &       31.486582 &  0.023252 &    82.722662 &  76.398430 &  3.819275 &            ELL \\
  2031113506311851904 & 105414150 & 13.237762 &       22 &       41.551937 &  0.057142 &    35.908104 & 100.607380 &  3.770568 &            ELL \\
  1996704839648530816 & 323630694 &  9.797777 &       17 &       38.942818 &  0.046335 &     7.545741 & 168.690275 &  3.741341 &            ELL \\
   251157906379754496 & 428692810 & 12.054820 &       21 &       49.165646 &  0.088727 &   296.675772 &  49.153139 &  3.607800 &                \\
  3331748140308820352 & 438063379 & 12.716515 &       13 &       41.455563 &  0.066978 &   225.268159 &  53.484464 &  3.547392 &                \\
  5963629779180627968 & 123810745 & 11.818235 &       13 &       25.285658 &  0.029502 &    11.368877 & 139.363823 &  3.184607 &        ELL, EA \\
  2929565719083290240 &  81733775 & 12.877993 &       28 &       49.930740 &  0.033133 &    32.473360 &  97.134994 &  3.078891 &        ELL, EB \\
  5596083868330026624 & 152577827 & 11.313450 &       30 &       29.111685 &  0.016003 &    57.846818 &  78.531008 &  2.901961 &             HI \\
  5545731836408967040 & 419154174 & 12.634378 &       23 &       24.275116 &  0.013377 &   983.518197 &  29.538204 &  2.625871 &                \\
  3123854268434443648 & 234801497 & 12.635286 &       10 &       46.279648 &  0.032723 &    87.885426 &  64.503233 &  2.440163 &            ELL \\
  2170359438392059776 & 267479007 & 13.284603 &       28 &       39.155950 &  0.066104 &    49.274869 &  77.844395 &  2.392820 &                \\
  2208323856913151360 &  12871969 & 12.969639 &       21 &       27.721788 &  0.062588 &    84.991016 &  64.424862 &  2.341165 &                \\
   440218930776209664 & 116842332 & 13.042632 &       25 &       27.773575 &  0.182453 &    47.537018 &  78.616544 &  2.274962 &                \\
  2004074312033531264 & 389024325 & 12.771193 &       18 &       32.358902 &  0.431375 &  1117.916540 &  29.162925 &  2.109731 &                \\
  5866096676853138432 & 330567799 & 11.415991 &       32 &       23.974285 &  0.083211 &    76.373014 &  63.171280 &  1.974368 &                \\
  2204345647749753216 & 411042914 & 12.717619 &       25 &       28.429789 &  0.145334 &    41.268379 &  77.860137 &  1.954839 &        ELL, HI \\
  3401898906303342848 &   6240885 & 12.385528 &       20 &       24.790207 &  0.142666 &    35.665994 &  78.726311 &  1.748532 &   ELL, ROT, HI \\
  5877107564251430400 &  45246020 & 13.712082 &       25 &       28.775150 &  0.229085 &   925.166866 &  26.935992 &  1.728053 &                \\
   427039394137929216 & 420024388 & 11.398898 &       25 &       24.639140 &  0.078857 &     5.436449 & 145.160304 &  1.707068 &       ELL, VAR \\
  5694373091078326784 & 155728890 & 11.913286 &       24 &       30.750462 &  0.017614 &    12.884807 & 107.469851 &  1.656499 &           EB22 \\
  4514813786980451840 & 451659136 & 12.565736 &       18 &       24.562475 &  0.136061 &    22.020434 &  88.831086 &  1.555261 &           EB22 \\
  5325390221595083520 & 400862980 & 13.442926 &       21 &       23.994865 &  0.338450 &   381.859396 &  36.053700 &  1.545096 &                \\
  5529029223988702976 & 181003751 & 10.852627 &       20 &       39.660534 &  0.096665 &   141.361843 &  47.405253 &  1.538690 &                \\
  4657646364721058304 & 404851003 & 10.861196 &       13 &       24.140335 &  0.052787 &   272.685820 &  37.752778 &  1.514074 &                \\
   948585824160038912 &   9961832 & 11.956474 &       10 &       21.798958 &  0.077334 &     8.201945 & 120.352267 &  1.468230 &      ELL, EB22 \\
  2219809419798508544 & 372337574 & 12.480452 &       17 &       38.396572 &  0.074999 &    10.865285 & 104.787932 &  1.284466 & ELL, ROT, EB22 \\
  5536105058044762240 & 123110322 & 12.154784 &       11 &       24.045340 &  0.085800 &    12.176559 &  98.074715 &  1.177076 &      ELL, EB22 \\
  4239611700214875904 &  95324628 & 11.914187 &       17 &       23.599092 &  0.404139 &    56.124101 &  62.948652 &  1.110104 &                \\
  4341334461557059328 & 157778582 & 13.509942 &       18 &       21.707628 &  0.379624 &   353.710453 &  33.458745 &  1.086982 &                \\
  3107879743172451456 & 177622520 & 11.904951 &       10 &       21.314194 &  0.026828 &    27.304940 &  72.539931 &  1.078762 &                \\
  5610703902634458112 &  63361138 & 11.006958 &       37 &       48.140770 &  0.021076 &  1003.250050 &  21.017484 &  0.964460 &                \\
  4295254307232999296 & 135660189 & 11.553295 &       16 &       29.559841 &  0.046062 &    10.832797 &  93.256394 &  0.907434 &        ELL, EB \\
   209136938490844928 & 369256285 & 12.401679 &       21 &       30.389870 &  0.042916 &    14.291517 &  83.789328 &  0.868692 &        ELL, HI \\
  6551893071549223040 & 175399204 & 11.893268 &       12 &       36.334988 &  0.092529 &    79.395211 &  46.970404 &  0.841566 &                \\
  5558841347897409408 & 130374599 & 11.852222 &       27 &       49.228947 &  0.067814 &    12.819832 &  84.129925 &  0.785513 &            ELL \\
  6609971199873698944 &  47505992 & 11.518994 &       12 &       36.472637 &  0.193909 &    18.342281 &  75.658301 &  0.777103 &                \\
  5337295389939374592 & 466883848 & 13.003701 &       14 &       27.571960 &  0.013766 &   274.397225 &  29.961135 &  0.764457 &                \\
  4278307358382586880 & 168891318 & 12.534010 &       10 &       29.479143 &  0.174147 &    11.962745 &  84.255175 &  0.707914 &                \\
  5461264982333157888 &  71585696 & 12.305203 &       55 &       21.673530 &  0.316210 &   762.774486 &  21.407609 &  0.662056 &                \\
  4039283116865495424 &  58882119 & 11.994610 &       15 &       20.938140 &  0.127662 &    79.548781 &  43.365410 &  0.655816 &                \\
  6475655404885617920 & 100475949 & 12.372097 &       15 &       20.482758 &  0.101280 &    46.094713 &  51.414310 &  0.639159 &                \\
  4345739070777246080 & 135817331 & 12.287071 &       13 &       29.691814 &  0.095857 &     6.195072 &  97.469449 &  0.586220 &       ELL, ROT \\
  5834976099807482752 & 428137933 & 12.196315 &       22 &       21.362688 &  0.046363 &    39.265479 &  52.202597 &  0.576909 &                \\
   421738201900868608 & 449984714 & 11.459711 &       27 &       33.416725 &  0.037187 &    16.552523 &  68.819819 &  0.557860 &            ELL \\
  2220072241734878720 & 372137170 & 11.751909 &       14 &       21.381530 &  0.179089 &   331.191043 &  24.850305 &  0.501488 &                \\
\bottomrule
\end{tabular}

        \begin{tablenotes}
                \item The remark EB22 means the system was included in \citet{El-Badry2022}.
        \end{tablenotes}
    \end{threeparttable}
\end{table*}

After the vetting in $\S$\ref{sec:lcs}, we are left with the \nVetted{} candidates in Tables \ref{tab:candidate_tableAB} and \ref{tab:candidate_tableC}. While some of these may represent good targets for spectroscopic follow-up to identify non-interacting compact objects, further analysis of the spectral energy distributions (including the low resolution \textit{Gaia} $BP/RP$ spectra) and their positions on the color-magnitude diagram (CMD) should be used to search for signs of a luminous companion and identify the best candidates for follow-up observations.

\begin{figure*}
    \centering
    \includegraphics[width=0.7\linewidth]{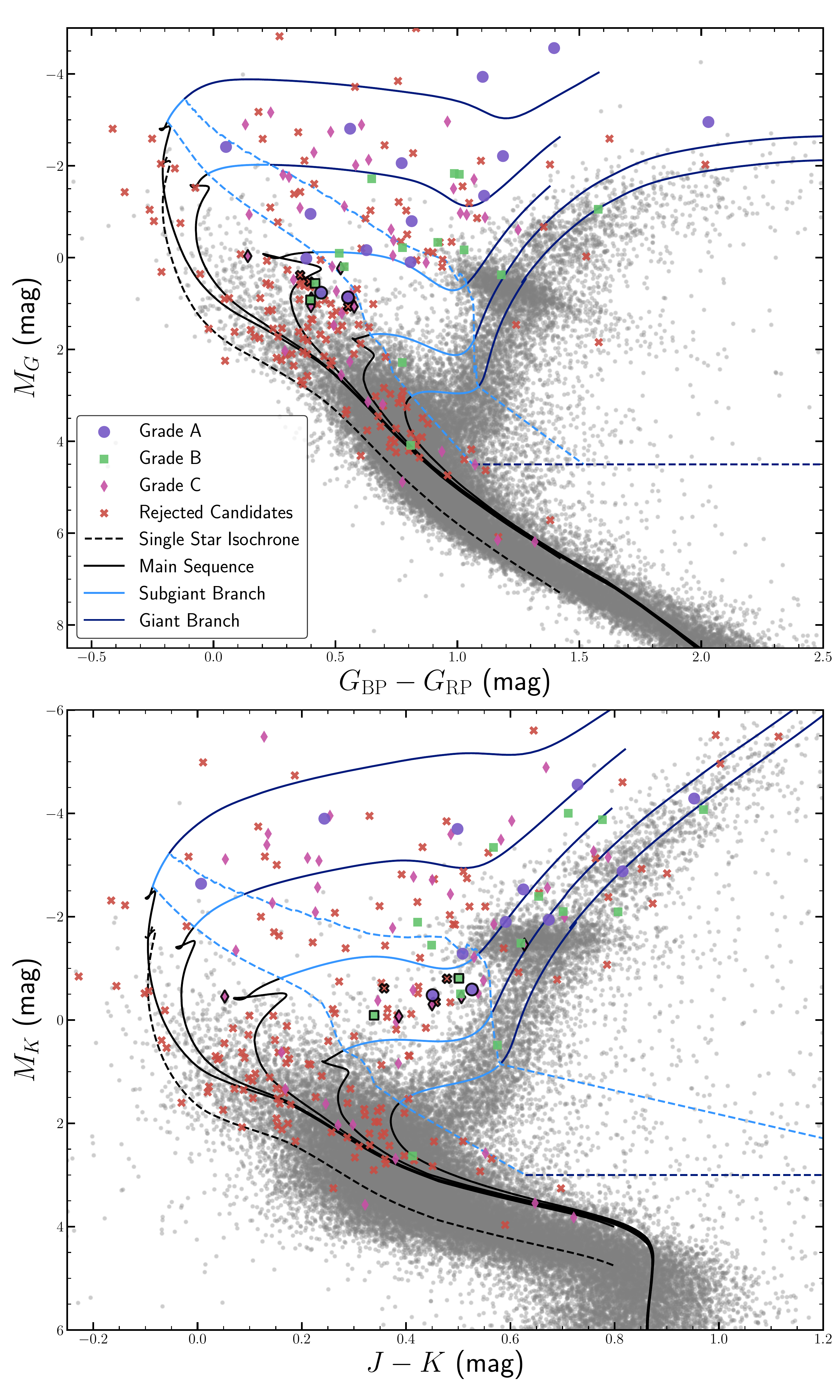}
    \caption{\textit{Gaia} (top) and 2MASS (bottom) color-magnitude diagram for the high mass function SB1 systems. The gray background shows a random selection of Gaia targets. The colored points show the Grade A, B, and C systems that remain after light curve inspection. The black crosses mark systems that were rejected based on the significance cut ($S<20$) or the light curves. Isochrones for equal mass binaries are shown as the solid lines, and divisions of the CMD from \citet{Rowan2022}. The black dashed line shows a single star isochrone.}
    \label{fig:cmd}
\end{figure*}

Figure \ref{fig:cmd} shows the Grade A/B/C targets on \textit{Gaia} and 2MASS CMDs after correcting for extinction. Solar metallicity isochrones where the flux in each band has been doubled to represent an equal mass binary are also shown. The divisions into evolutionary states shown by the colored dashed lines are described in $\S2.1$. Of the \nVetted{} candidates in Tables \ref{tab:candidate_tableAB} and \ref{tab:candidate_tableC}, \nVettedMS{}, \nVettedSG{} and \nVettedRG{} are main sequence stars, sub giants and red giants in the \textit{Gaia} DR3 CMD, respectively. In the 2MASS CMD, there were \nVettedMStM{}, \nVettedSGtM{} and \nVettedRGtM{} main sequence stars, sub giants and red giants, respectively. Of the \nVetted{} vetted candidates, only 51 had matching evolutionary states in both the \textit{Gaia} and 2MASS CMDs.


On the \textit{Gaia} CMD, the majority of the vetted systems are found in the Hertzsprung gap where few stars should dwell. This likely indicates the presence of a luminous companion instead of a dark compact object, where the intermediate $G_{\rm BP}-G_{\rm RP}$ colors are created by blending the spectral energy distributions (SEDs) of a cooler/redder star and a hotter/bluer star. When the same systems are examined on the 2MASS CMD, most of these systems now lie closer to the red giant branch, which supports the idea that a large number of these vetted systems are stellar binaries.
Indeed, \nElBadryTotal{} of our targets were also investigated by \citet{El-Badry2022} who argue that these are mass-transfer binaries consisting of a stripped giant with $M < 0.5$~M$_\odot$ and a blue companion. We highlight the systems that are in the \citet{El-Badry2022} catalog with black outlines on Figure \ref{fig:cmd} and label them in Table \ref{tab:candidate_tableAB}. 


\begin{figure*}
    \centering
    \includegraphics[width=\linewidth]{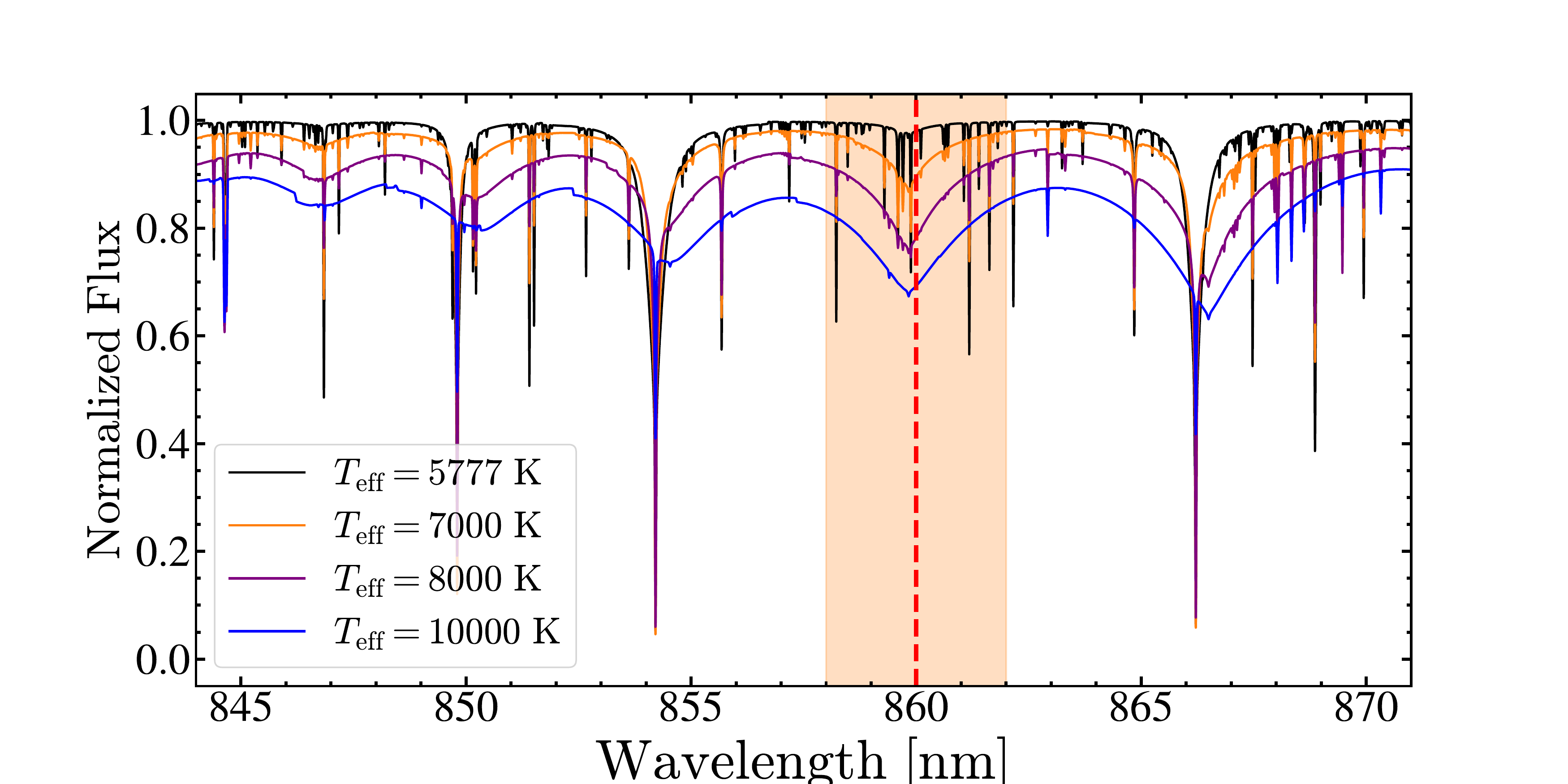}
    \caption{Synthetic spectra in the wavelength range covered by the \textit{Gaia} RVS spectrometer for main sequence stars with various $T_{\rm eff}$. The red dashed line and the orange shaded region shows the position of the H I absorption feature that is sensitive to $T_{\rm eff}$. }
    \label{fig:rvs_spectra_synth}
\end{figure*}

\begin{figure*}
    \centering
    \includegraphics[width=\linewidth]{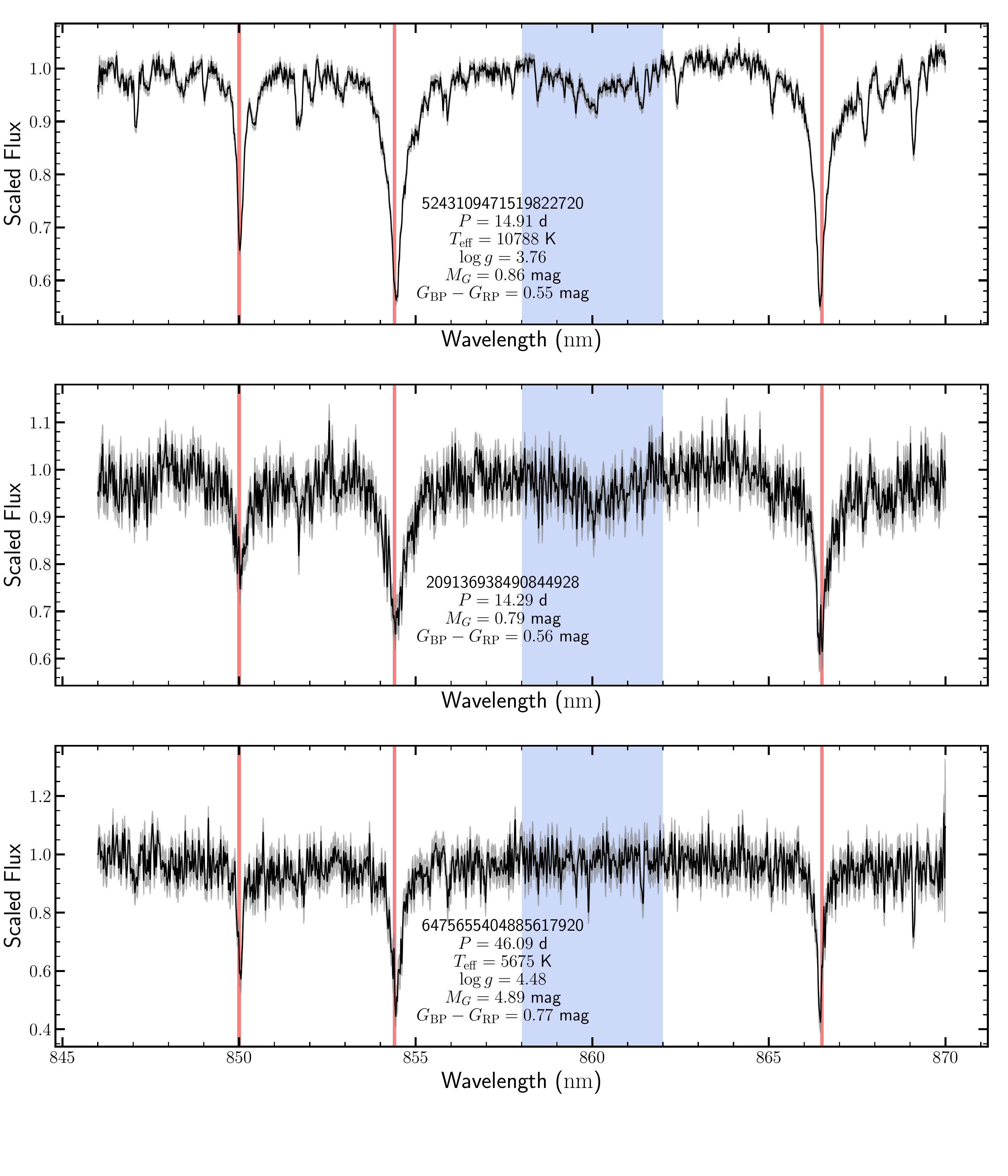}
    \caption{Mean Gaia RVS spectra for three main sequence targets. The top panel shows 5243109471519822720, a mass transfer system identified in \citet{El-Badry2022}. The middle panel shows one of our candidates with a similar spectrum, suggesting that it is also a binary with a hot accretor and cool donor star. The bottom shows a candidate with a spectrum that is more consistent with what is expected from a single star. In each panel, the \textit{Gaia} Source ID, period, and extinction-corrected absolute magnitude and color are shown. For the first and third targets, estimates of the effective temperature and surface gravity are given from \textit{Gaia}-Flame. In each panel, the red lines show Ca I lines, and the blue shaded region shows a H I line that is sensitive to $T_{\rm eff}$.}
    \label{fig:rvs_spectra}
\end{figure*}

Although the individual spectra are not available for each RV epoch, mean \textit{Gaia} RVS spectra with $R{\sim}11,500$ are available for \nRVS{} of the \nTotal{} targets (${\sim}28\%$). While the RVS spectra only span a limited range in wavelength (${\sim}845-871$~nm), they can be used to identify double-lined spectroscopic binaries. In particular, \citet{ElBadry2022} point out that the H I absorption feature in the RVS spectra at 860~nm is sensitive to the effective temperature. It is only present for $T_{eff}\gtrsim 7000$~K and then has an equivalent width that increases with temperature.  Figure \ref{fig:rvs_spectra} shows RVS spectra for three targets on our vetted list. The top panel shows 5243109471519822720, which is one of potential the mass-transfer systems from \citet{ElBadry2022} (their Figure 2) and the broad H~I feature lies in the shaded band. We examined the 11 available RVS spectra for the 39 targets classified as red giants in the 2MASS CMD (Figure \ref{fig:cmd}). We found that 6 of the 11 targets had evidence of this feature and so are likely stellar binaries with a red giant and a hotter stellar component. These systems are flagged with an `HI' in Tables \ref{tab:candidate_tableAB} and \ref{tab:candidate_tableC}.

\begin{figure*}
    \centering
    \includegraphics[width=\linewidth]{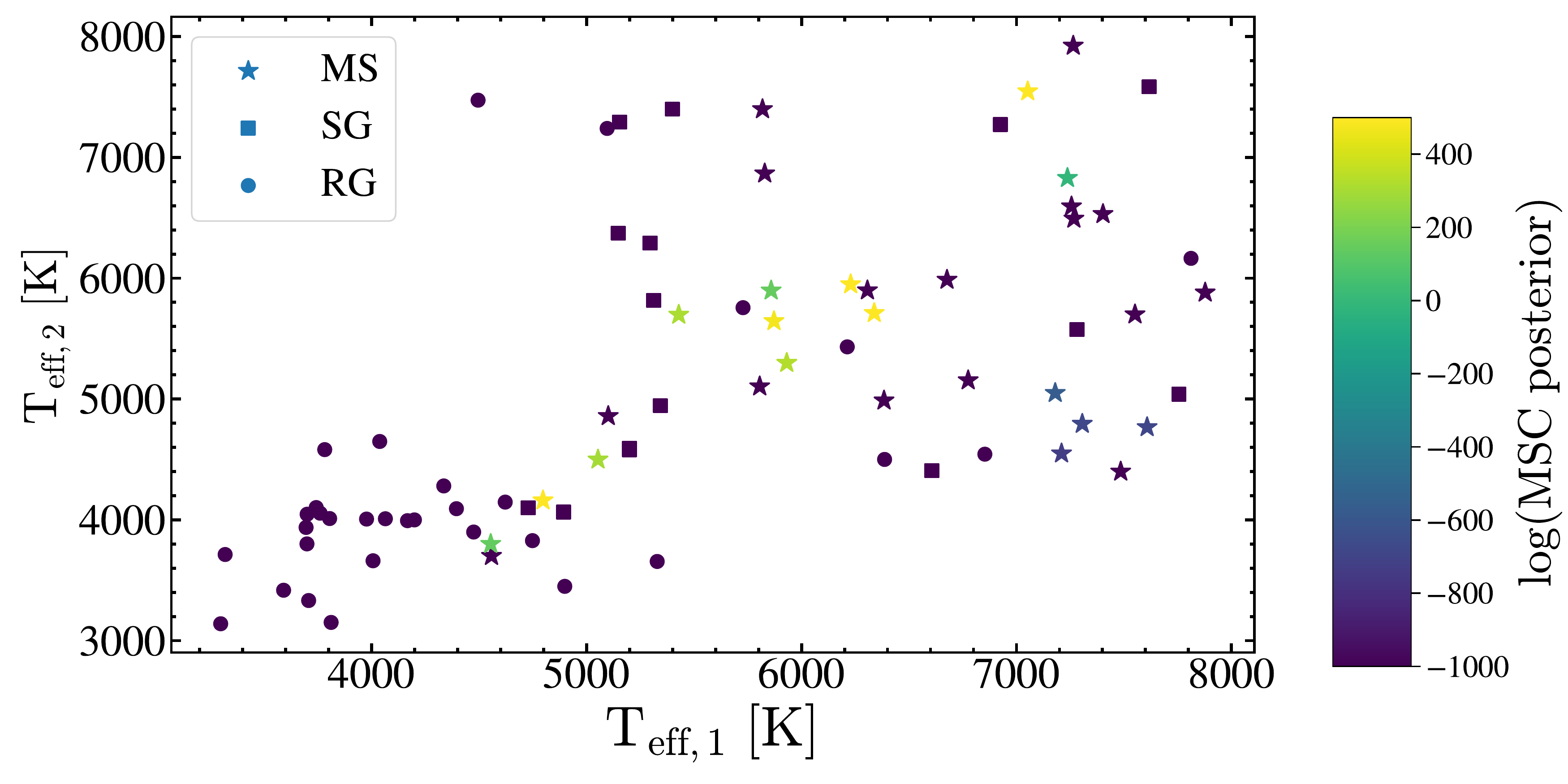}
    \caption{Comparison of the reported $T_{\rm eff}$ (primary and secondary) in the \textit{Gaia} DR3 APSIS/MSC module for the \nVetted{} candidates in Tables \ref{tab:candidate_tableAB} and \ref{tab:candidate_tableC}. }
    \label{fig:msc_teff}
\end{figure*}

The Gaia DR3 APSIS/MSC module (\citealt{Fouesneau2022}) models each system's $BP/RP$ spectrum as two stars in order
to search for unresolved binaries. This seemed an obvious way to identify stellar binaries with very different temperatures.  Unfortunately, the solutions are almost all reported to have very low posterior probabilities. Figure \ref{fig:msc_teff} shows the two MSC temperatures coded by the posterior probabilities. \citet{Fouesneau2022} note that the interpretation of the posteriors is presently not fully understood, but in comparisons to known GALAH unresolved binaries, the temperature differences rise from 135~K (258~K) for the primary (secondary) for the stars in the top 5\% of the posterior distribution, to 387~K (632~K) for those in the bottom 5\%. Bear in mind that if the systems in the Hertzsprung gap are due to combining a giant and a MS star, the temperature differences should be very large. Of the 79 vetted candidates, there were 24 that had $\Delta T_{\rm eff}>1000$~K. Of the 24 systems with $\Delta T_{\rm eff}>1000$~K, 13 were identified as sub giants or red giants in the \textit{Gaia} CMD (Figure \ref{fig:cmd}). This also suggests that some of these systems are in fact stellar binaries with a blended SED.

\section{Conclusions} \label{section:summary}

\textit{Gaia} DR3 provides $>181,000$ radial velocity solutions for single-lined spectroscopic binaries. We selected \nTotal{} SB1s with binary mass functions $f(M)>1.5~M_\odot$ (evolved) or $f(M)>0.5~M_\odot$ (main sequence) and identified \nGraded{} systems with good RV solutions ($S>20$) in \textit{Gaia} DR3. Using light curves from ASAS-SN and TESS, we vet the selected SB1s to remove false positives from eclipsing binaries. We identified \nEBsTotal{} eclipsing binaries and \nELLsTotal{} ellipsoidal variables. We produced a catalog of \nVetted{} vetted SB1 candidates. The positions of the vetted SB1 candidates on the \textit{Gaia} and 2MASS CMDs suggest that many, if not all, of these systems are binaries with luminous companions. We do not find any strong candidates for non-interacting compact object+star binaries with just the \textit{Gaia}, ASAS-SN and \textit{TESS} data. Further spectroscopic follow-up is necessary to determine whether these systems are non-interacting compact object binaries.


After noting some discrepancies between the \text{Gaia} periods for the eclipsing binaries and ellipsoidal variables we identified among the candidates, we did a broader comparison using the 596 detached eclipsing binaries from \cite{Rowan2022}
and the 311 spectroscopic SB9 (\citealt{Pourbaix2004}) binaries that had SB1 solutions. 
For the eclipsing binaries, ${\sim}89\%$ had periods that agreed to within $\pm10\%$ and ${\sim}40\%$ had eccentricities that agreed to within 0.05. For the spectroscopic binaries, we found that ${\sim}85\%$, ${\sim}80\%$ of these systems had periods and semi-amplitudes that agreed to within $\pm10\%$. However, only ${\sim}70\%$ of these systems had eccentricities that agreed to within 0.05. 

\section*{Acknowledgements}

We thank Las Cumbres Observatory and its staff for their continued support of ASAS-SN. ASAS-SN is funded in part by the Gordon and Betty Moore Foundation through grants GBMF5490 and GBMF10501 to the Ohio State University, and also funded in part by the Alfred P. Sloan Foundation grant G-2021-14192.

This work presents results from the European Space Agency space mission Gaia. Gaia data are being processed by the Gaia Data Processing and Analysis Consortium (DPAC). Funding for the DPAC is provided by national institutions, in particular the institutions participating in the Gaia MultiLateral Agreement.

This paper includes data collected with the \textit{TESS} mission, obtained from the MAST data archive at the Space Telescope Science Institute (STScI). Funding for the TESS mission is provided by the NASA Explorer Program. STScI is operated by the Association of Universities for Research in Astronomy, Inc., under NASA contract NAS 5-26555.CSK, KZS and DMR TESS research is supported by NASA grant 80NSSC22K0128.

TJ, KZS and CSK are supported by NSF grants AST-1814440 and 
AST-1908570. TJ acknowledges support from the Ohio State Presidential Fellowship. TAT is supported in part by NASA grant 80NSSC20K0531. TAT acknowledges previous support from Scialog Scholar grant 24216 from the Research Corporation, from which this effort germinated.

\section*{Data Availability}

The Gaia DR3 data, the ASAS-SN and TESS light curves are all publicly available.


\clearpage
\bibliographystyle{mnras}
\bibliography{ref_dr3bhs} 


\bsp	
\label{lastpage}
\end{document}